\def\year{2021}\relax
\documentclass[letterpaper]{article} 
\usepackage{aaai21}  
\usepackage{times}  
\usepackage{helvet} 
\usepackage{courier}  
\usepackage[hyphens]{url}  
\usepackage{graphicx} 
\usepackage{natbib}  
\usepackage{caption} 
\usepackage{xcolor}
\definecolor{malachite}{rgb}{0.2, 0.6, 0.2}
\definecolor{mango}{rgb}{.9, 0.6, 0.3}

\title{Open Player Modeling: Empowering Players through Data Transparency}
\author {
    Jichen Zhu,\textsuperscript{\rm 1*}
    Magy Seif El-Nasr \textsuperscript{\rm 2*}\footnote{Both authors contributed equally to the paper.}\\
}
\affiliations {
    \textsuperscript{\rm 1} IT University of Copenhagen \\
    \textsuperscript{\rm 2} UC Santa Cruz \\
    jichen.zhu@gmail.com, mseifeln@ucsc.edu \\
}

\begin{document}
\maketitle

\begin{abstract}

Data is becoming an important central point for making design decisions for most software. Game development is not an exception. As data-driven methods and systems start to populate these environments, a good question is: can we make models developed from this data transparent to users? In this paper, we synthesize existing work from the {\em Intelligent User Interface} and {\em Learning Science} research communities, where they started to investigate the potential of making such data and models available to users. We then present a new area exploring this question, which we call {\em Open Player Modeling}, as an emerging research area. We define the design space of Open Player Models and present exciting open problems that the games research community can explore. We conclude the paper with a case study and discuss the potential value of this approach. 

\end{abstract}



\section{Introduction}

As data science and Machine Learning (ML) become increasingly adopted in a wide range of everyday digital products, there is a rapidly growing demand to make the underlying algorithmic processes and models more transparent. As a response, new research areas such as eXplainable AI (XAI) and Interactive Machine Learning (IML) have emerged to investigate how to make data science and ML models less opaque and more interpretable~\cite{fails2003interactive,Gunning2017,amershi2014power}. Researchers have made progress to open the black box of various ML algorithms in a wide range of domains, including image recognition, Natural Language Processing (NLP), and personalized apps. However, how to apply such progress to make these models and processes open to end-users, who may not AI experts, remains an open  problem~\cite{stumpf2009interacting,kulesza2013too}.   

With a few exceptions\cite{Zhu2018,hooshyar2020transparent,zhu2020player}, current games research on how to improve the transparency of AI and data is limited. This paper focuses on player modeling, an active game AI research area that studies computational models of players in games~\cite{yannakakis2013player}. Player modeling can benefit from increased transparency because it is often used by different human stakeholders. In the game development process, analysts use player models to gain further insights into different player types and improve the gameplay~\cite{el2016game}. Game designers use them for personalization, such as tailoring game content to individual players' learning needs~\cite{Kantharaju2018AIIDE}. Players are beginning to use player modeling. For example, an e-Sports player can use automatic coaching systems that model her gameplay and recommend different gameplay strategies tuned to her specific skills and preferences. These applications require player models to be transparent to not only experts in interpreting behavioral data but also regular players. Finally, as one of the desiderata for human-centered AI~\cite{Miller2019}, transparency can contribute to the fairness of player models and engender player trust.  




%
This paper proposes a new research area --- {\em Open Player Modeling}(OPM) --- to make computational models of players accessible and more transparent to {\em  players themselves}. This involves answering questions such as WHAT data or models would be useful for players, HOW would such data be communicated, and WHEN to show these models or the resulting processes, such as recommendations or predictions, to players in the gameplay experience. 
%
%
Our core position is that open player modeling is a promising research area that can potentially empower players to learn from their own data, increase their trust in AI, and enhance the effectiveness of some games as interventions. 
In e-Sports, OPM can be used to enhance the experience of spectators and to help players learn from other players' gameplay strategies. 
In this paper, we illustrate the usefulness of OPMs through a case study showing how an OPM can be applied to an educational game on parallel programming. We argue that OPM can increase players' reflection on their learning strategy and self-regulate learning more effectively. 

The contribution of this paper is the following. 1) We identify Open Player Modeling as a promising new research area by synthesizing related research in separate communities: games, intelligent user interfaces, and intelligent tutoring systems. 2) We provide a theoretical framework to map out the design/problem space of open player modeling and identify key open problems. 3) We offer a case study of how OPM can be used to enhance players' experience and capabilities. 


\section{Related Work}
Constructing computational models of how users interact with digital systems is the subject of user modeling research\cite{fischer2001user} and its domain-specific forms such as player modeling\cite{yannakakis2013player} and student/learner modeling\cite{vanlehn1988student}. These models are then commonly used in many areas for data mining, data analytics, and automatic content adaptation. 
Currently, the vast majority of these models are hidden from humans or designed for trained analysts~\cite{nguyen2015glyph}. In this section, we discuss existing work that attempts to make them more open to end users such as players. 


\subsection{Open User and Learner Modeling}
Most existing research efforts on showing users computational models of their behavior concentrate in the Intelligent User Interfaces (IUI) community. Most notably, {\em Open User Modeling} (OUM)\cite{Brusilovsky2011} aimed to make the process and products of user models explicit and available to users themselves. Existing research found that OUMs can improve a user's understanding of the system \cite{Knijnenburg2012,Gretarsson2010} and encourage reflection on users' data and actions \cite{bull1995extending,bull2004supporting,Gretarsson2010}. When applied to the context of web-based e-learning, studies showed that OUM, sometimes referred to as Open Learner Modeling (OLM), helped students improve self-reflection and the meta-cognitive processes \cite{Hsiao2013, Hsiao2017, Hsiao2012, Hsiao2012a, Law2017, Kump2012, Brusilovsky2011}. In addition to individual open models, recently, researchers started to explore the social aspects of OUMs. Motivated by social comparison theory \cite{suls1991social}, Hsiao and Brusilovsky \cite{Hsiao2017} proposed what they called Open Social Student Models by making a student's model visible to other students going through the same learning experience. In a user study, they observed that students ``spent more time on [the system], attempted more self-assessment quizzes, and explored more annotated examples'' (p. 10). 


Open Player Models (OPM) extend the main idea behind OUM/OLM to computer games. The majority of existing work on OUM/OLM is based on adaptive web applications where hyperlinks are the primary source of user interaction. As argued below, player interactions within games tend to be more complex than with web applications. Thus, 
OPMs face unique challenges in terms of how to open the models and how to make the resulting information useful to the players. 

\subsection{Open Models in Games}


Traditional player models, a component of the AI systems in games, are used directly by the game system to, for example, adapt the gameplay experience. They are typically not designed to be open to human users. As game user research develops, post-hoc player models start to be developed and are visualized for user researchers, game analysts, game business teams, and game developers. Such models are then used in the game production pipeline for various purposes, including enhancing the game design or tuning the technical aspects of the game \cite{el2016game}. Since these models are designed for highly trained experts, not the average players themselves, they are often highly technical.


The most developed applications of OPMs for non-technical stakeholders are dashboards of player performance. Dashboards are broadly used in modern computer games to display player performance stats, such as win rate, quality of solutions in puzzle games, and average speed in racing games. More recently, with the increased popularity of e-Sports, more sophisticated dashboards have been developed. Such dashboards have also been extended for spectators to show what is happening within the match \cite{koomen2020esports, kokkinakis2020dax, charleer2018real, kriglstein2020part}. 
Recently an increasing number of companies, e.g., SenpAI and Mobalytics, focus on developing AI-enhanced dashboards, such as recommendation systems targeting different kinds of MOBA games. 
%
%

On the research side, several works investigated dashboards for various games \cite{charleer2018real, kokkinakis2020dax, koomen2020esports, robinson2017all}. For example, Charleer et al. (\citeyear{charleer2018real}) sought to explore the effect of dashboards on spectators’ experience and how they were used to gain insight into the game. They ran surveys to study the value of different metrics to help design their dashboards. Based on this study, they developed specific dashboards geared towards the different games studied and evaluated their usefulness. They found that the dashboards were useful in that they facilitated analysis and interpretation, but some participants wanted access to data as they liked to dig into the data themselves and took pride in doing so. Further, researchers discussed issues of trust and warned about the complexity of the visualizations as well as issues of cognitive load.
%
Most relevant is a proposed idea called Transparent Player Models~\cite{hooshyar2020transparent}. In their proposal, researchers adapted Open Learner Models into educational games and proposed to use learning analytics and visualization to expose learner models to players. While similar to \cite{hooshyar2020transparent}, our work encompasses a broader scope of player modeling, in addition to educational games, we include MOBA games, health games, and other serious games. Further, our work is situated in games research, specifically looking at player modeling as part of the player experience and questioning how players can benefit from viewing models of their gameplay data. 


\section{Open Player Modeling} 
\begin{figure}
    \centering
    \includegraphics[width=\columnwidth]{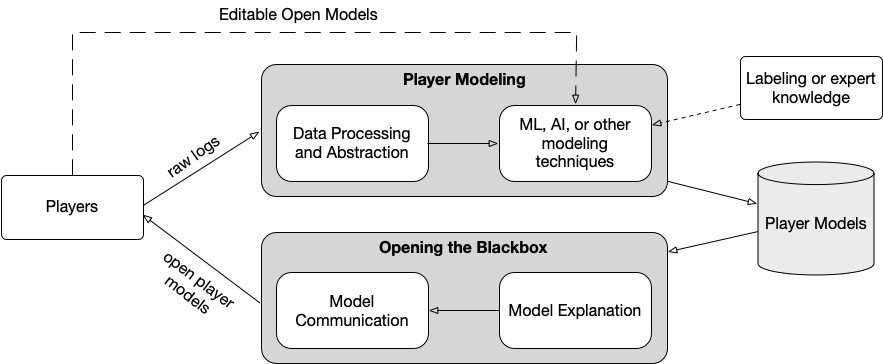}
    \caption{Components of Open Player Modeling}
    \label{fig:OPM}
\end{figure}

We propose the new term {\em Open Player Modeling (OPM)} to unify the range of practices of showing players computational representations of their gameplay data. These models can vary in their computational complexity --- ranging from players' in-game performance through descriptive statistics (e.g., average speed) to cognitive, affective, or behavioral models of player characteristics (e.g., learning preferences in educational games). 
Figure \ref{fig:OPM} shows the key components of OPM. First, player's gameplay log data is processed to produce a player model. In some cases, human experts are involved to provide labeling and domain knowledge to the modeling process. Next, these models are explained and communicated to the players and other stakeholders (e.g., game analysts) through text, visualization, or other media. In cases of complex player models with low interpretability, such as deep neural network-based player models, extra effort is needed in producing explanations to the underlying models before they can be communicated to the players.


\subsection{Open Player Modeling as a New Research Area}
As shown above, there are pockets of knowledge about open models in different research communities. So is OPM simply an extension of the prior work to a new domain, or is it fundamentally different from them? Below, we argue that there are four reasons why OPM is the latter.

First, player data is typically more complex than the type of user data applied to OLM and OUM research. As summarized above, most existing OLM and OUM research are conducted based on how users interact with hypermedia applications. In comparison, players interact with games with higher frequency and longer duration. A 2020 international  report\footnote{\url{https://www.limelight.com/resources/white-paper/state-of-online-gaming-2019/\#spend}, last accessed on January 31, 2021.} shows that although gamers typically play for an average of 1.37 hours at a time, many of them play for longer sessions. In the U.S. players' average longest consecutive play sessions are 5.10 hours. 
All the above statistics indicate the size and complexity of player data. 

Second, player modeling techniques are different from those used in other open models domains. Because play experience often involves time and virtual space, most player models involve a way to deal with the time and space continuum. Some models aggregate over these dimensions \cite{drachen2014comparison}, while others specifically look into sequence analysis methods that span time and space (e.g., \cite{sifa2018profiling, kleinman2020and, pfau2018towards, min2014deep}). 

Third, most player models will integrate some way of measuring or integrating preference, emotions, and engagement (some examples of these models have been discussed in Yanakakis et al.'s paper and book (\citeyear{yannakakis2013player, yannakakis2018artificial})). These factors are important for personalizing the play experience and also for developing game content. Due to the influence of game characters on team performance within a game like DoTA or League of Legends, such aspects will also need to be modeled (e.g., \cite{chen2018art}). Such types of models have been exemplified in many of the MOBA AI types of recommendation systems, including in Senpai.gg and Mobalytics.gg, which recommend items, characters, or other aspects of gameplay that are based on player state modeling.

Fourth, how players interact with OPM may be different from other open models, imposing new research questions for open models in the context of play. Players' desire to improve their in-game performance and win the game may provide them with the extra motivation to use OPM. 
Compared to users of other open models, gamers are more familiar with viewing information about their activities. Many games already use player stats to communicate how well each player is doing. In some complex games, players have to make sense of a wide range of stats and use them to inform gameplay decisions. Having this background means players may be more prepared to make use of more sophisticated open models in ways they may not be outside games. It thus makes OPM an exceptionally suitable domain to investigate eXplainable AI problems of communicating complex AI decisions to end-users.

\subsection{Design Space of Open Player Modeling}
This section maps the problem space of OPM along two main dimensions: the nature of the player models and their level of openness. 

\subsubsection{I. Types of Player Models}

There are different ways that different models in the literature have been categorized. For example, Yanakakis et al. (\citeyear{yannakakis2013player}) categorized computational player models in terms of their computational structure defining them as model-based techniques (techniques that embed a theoretical or expert model that is used to derive the player model), model-free techniques (those that construct a model based on data without a specific knowledge or theory), and hybrids that mix model-based and model-free techniques. Another approach to categorize models is the approach borrowed from statistics or engineering, where models are categorized into: (a) descriptive, where data is used to show or describe what has happened, this can be done at different levels of abstraction, (b) predictive, where data is used to predict an outcome or a state, (c) diagnostic, which extends a descriptive model to specifically answer specific questions, and (d) prescriptive extends predictive models to prescribe what to do next. 

While these frameworks are useful conceptual tools from the algorithmic perspective, they contain information that players may find unnecessarily complicated. Since OPM is used by players, we propose an adapted framework to classify player models from the players' perspective. 
{\begin{enumerate}
      \item \textbf{Descriptive Models} tend to describe a system or a phenomenon, constraints, complexity, and interaction. An example is a descriptive model that contains a player's problem-solving process using the gameplay sequence of how she solves a puzzle in a level (e.g., \cite{ahmad2019modeling, kleinman2020and}). 
      Models can vary in terms of the abstraction level that they encompass and they can offer players information about what has happened. 
    \item \textbf{Prediction Models} offer a forecast on a state of the game or the player \cite{yannakakis2013player, camilleri2017towards, henderson2020multimodal, shaker2013fusing}. Some of these models may target the prediction of an outcome, e.g., skill level or win/loss \cite{kang2020sports, lan2018player, norouzzadeh2017predicting}. Other models may predict players' state emotions \cite{yannakakis2013player}, churn \cite{castro2015churn, tamassia2016predicting}, or players' next action or a goal based on observed behavior \cite{min2014deep, harrison2014survey, harrison2011using}. Players can use these models to get a glimpse of the likely outcomes in the future and assess their current gameplay accordingly. 
    \item \textbf{Reflection Models} offer players information about their play, such as score or strategy, as a way to allow players to reflect on their performance. 
  
\end{enumerate}}

An alternative way to categorize player models is based on the type of service they offer. For example, for reminders, these models tend to be context-aware and use expert knowledge to show specific information to players based on their play and the data analyzed. A good example of this model can be seen in the Senpai.gg tool, where players can use the reminders to improve their gameplay. Another service type is recommendations. 
These models offer recommendations for items, characters, or specific elements to player pre- or during play \cite{chen2018q, chen2018art, bertens2018machine}. 

\subsubsection{II. Level of Openness}
We classify OPMs into the following levels of openness based on how much of player modeling process is revealed to the players and whether players can modify their models.

\begin{enumerate}
    \item \textbf{Open Model Outcome} has the lowest level of openness. It communicates to players the {\em result} of the player models, without detailed information of how the models were calculated from the players' gameplay behavior. For example, an OPM of an adaptive game similar to {\em Left4Dead} may communicate to the player that its model of her is that she is a highly-skilled player, and therefore the game is sending her more difficult enemies. Notice this type of OPM does not contain information about {\em how} the user model is computed. It focuses on {\em what} the model is. The main benefit of this type of OPMs is its simplicity. Especially when players are engaging in time-critical gameplay, such as combat, extra information of the model can be overwhelming.
    
    \item \textbf{Open Model Process} reveals the computational process through which the outcome of player models are calculated from the input of player behavior. For instance, this type of OPM can use visualization to display the clusters of players based on their gameplay and show a player where her gameplay is situated. It can also be used to explain certain predictions that the game AI has made (e.g., probability of winning a game). The main benefit of this type of OPMs is that they can help players understand WHY and HOW they are modeled in a particular way. 
    
    \item \textbf{Editable Open Models} affords the highest level of openness. In addition to making the player model open to the players, they allow players to modify their own models. When designed well, editable models can improve the accuracy of the models by allowing players to make corrections. For instance, editable OPMs can be used in adaptive games for physical rehabilitation patients. The game can model a player's joint limitations based on her movements in the gameplay and other information~\cite{pirovano2012self} in order to generate targeted exercises. However, if there is an error in the player model, it could likely lead to pain and additional injury. Through Editable Open Models, players can see their own models and make appropriate corrections so that the game can provide the most suitable exercises. 
    Since Editable Open Models allow players to directly modify the player models, there is a higher requirement for players to understand how the models function and therefore provide the correct editions. Additional design considerations are also needed to align players' interests (e.g., the desire to win) with the model accuracy. 
    
\end{enumerate}


\subsection{Open Problems}
This section discusses the main open problems of OPM. 

\subsubsection{Make Player Models Explainable and Transparent}
Opening player models to players requires the underlying player models to be explainable, especially for the models with higher levels of openness. The game AI community has developed a large body of work to model players, but the vast majority of them are black boxes. A key open problem for OPM research is how to improve the transparency of player modeling. For systems that use AI modeling techniques that have low interpretability (e.g., deep learning), how can we still offer insights to players about why the player model classified them in a certain way. For other models, how can we choose the appropriate level of abstraction to reduce the noise of low-level actions and highlight the main trends that are useful to differentiate among the different players?

\subsubsection{Communicate Open Models to Players}
Once we have an explainable player model, research is needed to effectively communicate the model to players. Currently, information visualization has been a primary tool to convey player stats. As OPMs may contain more complex information, more research is needed to expand the current work on visualization and multi-modal explanations (e.g., visualization, text, and voice). What is the information that the players want to know about their own models? How can we visualize high-dimensional player data and gameplay traces? For OPMs of open model outcome, how can we help players establish the connection between their gameplay actions and the model outcome? For OPMs of open model process, how can we provide relevant information without overwhelming the player? And for editable open models, can we provide sufficient scaffolding so that the players can provide more informed edits to the model?

\subsubsection{Design Player-AI/Model Interaction in the Eco-System of Games}
Compared with other eXplainable AI domains, a unique aspect of OPM is that it needs to be embedded in the rich game-player interaction context. This allows us to investigate new OPM and explainable AI research questions in a wide variety of contexts. For example, we can explore when and how to provide OPM information to players in the entire player experience. In open learner modeling literature, the models are often displayed at the end of a pedagogical module for reflection.
However, in educational games, it is possible that OPM can be used to provide guidance and encourage reflection {\em during} the gameplay. How can we address different player needs at various stages of the game and incorporate OPM accordingly? Can we even align the gameplay incentives with OPM so that players are motivated to provide accurate information to the editable player models? One notable design challenge is the potential conflict between OPM and immersion. OPM is not for every game. Directing players' attention into the player models, like any extra-diegetic features, may interrupt the narrative or emotional immersion. Game designers need to be mindful of these interruptions. Alternatively, OPM can be directly incorporated into the core game mechanics (e.g., as a meta-gaming feature\cite{Kleinman2020}). 


\subsection{Ethical Considerations}
By increasing the transparency of player models, OPMs will highlight the ethical issues that have underlined user modeling in games. Similar to the research of ethics of AI ~\cite{Hagendorff2020,Jobin2019} and user modeling~\cite{Mobasher2020}, making sure that OPMs embody human-centered values, such as fairness, accountability, and privacy, is a challenging research problem. For example, when we model players using emotions or webcam data, how would we open such models while also ensuring privacy. These issues need further examination. 
%
Another example is that game analytics tools regularly use aggregation (e.g., averages, typical gameplay paths, heatmaps) to abstract player data. While they are useful to describe ``average'' players, those who play differently can receive less design attention and thus be marginalized. In an OPM for learning (e.g., see case study below), we need to make sure that less popular problem-solving processes do not get buried and the OPM can serve all players. 

Another example of the ethical considerations of OPM is privacy. A good example is alluded to earlier on emotions or visual identifiable information embedded in the models. How do we share players' traces in a socially responsible way? By showing the different gameplay strategies, the OPM can reveal struggling players. Would it expose these players to potential toxic behaviors in games? Or, can we steer what OPM reveals towards self-improvement and assistance from the game and its community?  Finally, player modeling is rarely completely accurate. Opening these models up to players further raises questions around accountability. What happens when the model incorrectly characterizes a player? Editable open models allow players to identify and repair these instances, but they also impose additional challenges of explainability and preventing players from potentially exploiting the system.  







\section{Case Study}

Below we discuss a case study of OPM in an educational game. 
Note this work is still ongoing and the validation of this model has yet to be measured.

\subsection{Context: Learning Game called \textit{Parallel}}
{\em Parallel} is a single-player 2D puzzle game, designed to teach parallel and concurrent programming core concepts, especially {\em non-determinism}, {\em synchronization}, and {\em efficiency}, to Computer Science undergraduate students~\cite{zhu2019programming,ontanon2017designing}. In each level, arrows (the equivalent of threads) move along a pre-determined set of tracks (a program). A player places semaphores ({\em wait} function in parallel programming) and buttons ({\em signal}) in order to direct arrows to pick up packages and deliver them to the designated delivery points. In addition, the game contains directional switches to represent conditional statements. In essence, the player designs a {\em synchronization mechanism} to coordinate threads executing in parallel. Once the player successfully delivers all required packages, she wins and moves to the next level. 
Because the arrows move at random speeds in each simulation, the puzzles in {\em Parallel} are non-deterministic. The game allows learners to test their solutions with {\em one} configuration of arrow movement schedule at a time before they submit their final solution. At that point, the game will systematically check if the solution works in {\em all} configurations. 
%



From an in-class user study~\cite{zhu2020understanding}, we observed the difficulty for students to think in a multi-threaded way. Specifically, players tend to assume the same problem-solving strategy even when they encountered problems. It often leads to stagnation in learning. Based on this observation, we hypothesize that enabling players to see their own problem-solving patterns through OPM, especially in the context of others', can encourage reflection, motivate them to explore other problem-solving strategies in the game, and hence enhance learning.

\subsection{\textit{Open Player Modeling for Parallel}}
\subsubsection{Player Model Type}
We believe that the use of OPM here can potentially enable players to reflect on their own problem-solving patterns and explore new ways when comparing them with other players'. Here we decided to use a descriptive model, where low-level gameplay actions, such as mouse clicks and button presses, are abstracted into higher-level actions such as level test, placing a semaphore, linking a semaphore to a signal, etc., which focuses on decision-making and problem-solving strategies. A key challenge here is that, directly showing players' every gameplay action will be overwhelming in terms of the number of actions as well as the complexity of the action sequences. Yet the established technique of showing aggregated player stats hides necessary information about problem-solving strategies. 
This challenge creates a rich domain for OPM to explore how to model player action sequences and how to make them open and transparent to players for reflection~\cite{villareale2020reflection}.

\subsubsection{Level of Openness}
Since our goal is to improve players' reflection and learning, we intend our player model to be open in a way that players can use to compare their gameplay/problem-solving process with others'. We decided to explore Open Model Outcome as the foundation type of OPM. In our case, the model is the individual players' problem-solving processes, expressed as sequences of abstracted player actions during a level. 

\begin{figure}
  \centering
    \includegraphics[width=0.3\textwidth]{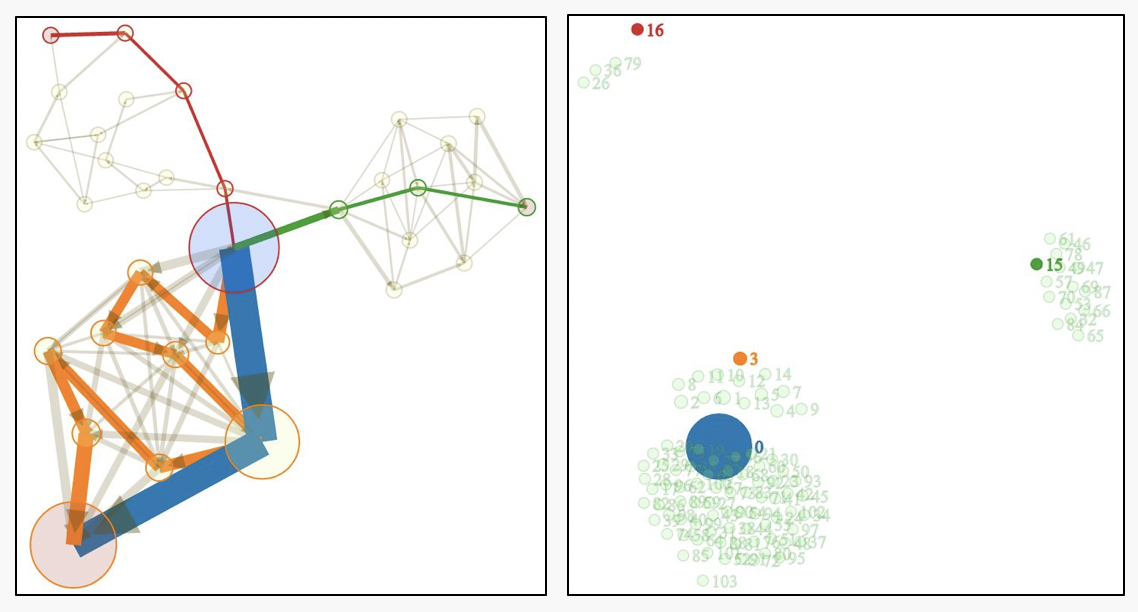}
  \caption{A screenshot from Glyph where a user highlighted four different sequences or patterns for solving this game level. These patterns are highlighted in red, blue, green, orange shown in the sequence graph (right) with the corresponding traces shown in the state graph (left). }
  \label{fig:Glyph2}
\end{figure}

We use visualization, through the \textit{Glyph} visualization system, to communicate the OPM to the players. 
\textit{Glyph} \cite{nguyen2015glyph}, developed to display traces of players' sequences of actions, supports comparison across many players. 
%
%
It is composed of a dual-view interface that shows data from two related perspectives: a state graph and a sequence graph. A state graph shows a sequence of abstract behaviors and abstracted states. As shown in Figure~\ref{fig:Glyph2} Left, this state graph representation is a node-link graph. The nodes represent different game states --- a game state is a context of the game where a player action is taken. The links between these nodes are actions that a player took to get from one state to another. 

To facilitate comparison of individual action sequences, we augmented the state graph with a synchronized sequence graph showing the popularity and similarity of sequence patterns exhibited by users (see Figure~\ref{fig:Glyph2} Right). Each node in the sequence graph represents a full play trace, the size is an indication of popularity, i.e. how many players made these actions. Further, the distance between each node provides a visual representation of similarity/dissimilarity. 
For this particular level, the {\em Glyph}-based OPM revealed three distinctive ways that players in our dataset have solved it.  The sequence graph (Figure \ref{fig:Glyph2} Right) shows three distinct clusters of the gameplay sequences. 
A player can locate her gameplay sequence and see how it relates to others.


To inspect this even further, the visualization system allows users to interact with the graphs, highlighting different sequences to inspect the decision-making steps in detail. Figure ~\ref{fig:Glyph2} shows such rendering of four different paths towards solutions for the level. This system then demonstrates an example Open Player Model - visualizing a descriptive process-oriented Model. One can thus imagine a player who may be struggling with a level to inspect the data from another player who was successful and follow through the sequence: ``added a wait statement for red arrow", ``tested and failed", ``adjusted the wait statement for the red arrow", ``tested and passed". Visualizing such patterns can give the player a way to inspect their own actions, reflect on them and also compare them to others. The specificity of the design and the visualization will depend on the model used.  

We are currently tackling several open problems. First, we are investigating whether {\em Glyph}-based visualizations are interpretable by regular players. It is possible that players may find the state graph too difficult to understand. Through user studies, we will test and refine appropriate techniques that are informative and understandable to our target player group. There needs to be a balance between the expressivity of the visualization and its understandability. 
Second, we are also exploring the type of interactions and features needed to make our OPM useful for reflection. For example, we anticipate using annotations to allow users to leave notes or reflections on detailed problem-solving processes. Further, we also plan to use AI techniques to guide attention to similar sequences or add filtering options. These interactions will need to be tested through user studies. 
Third, we also need to investigate the type of player model to construct for an effective OPM. This is due to the fact that OPMs are tightly coupled with how explainable the resulting visualization is to the players. As mentioned above, there is a tension between models that capture sufficient information about problem-solving patterns (tend to be complex player models) and models that can be easily understood by players (tend to be simple player models). 

\section{Discussion and Conclusion}
While the case study above specifically showcased one example in the space of possible OPMs, there are many other examples that one can imagine for {\em Parallel}. These include recommendation models where the system recommends actions or strategies to players based on their models and level. We can also imagine models that give players reminders about specific aspects of the level to pay attention to, again given players' models. These types of models are different from the descriptive models described above and can be effective for learning. However, they also need to be equipped with more explanations for the information presented. These could include different ways to establish the openness of the model. We can imagine, for example, a system that explains the process by which the model is developed, i.e. Open Model Process. We can also imagine that sometimes such models need to be edited or adjusted due to low accuracy or issues with the model, in such cases, designers can investigate the use of editable models. 
Researchers can develop different types of OPMs given the framework we outlined through adjusting the model's type and level of openness. 
We hope these dimensions can help establish and further expand the work on player modeling to deliver better game mechanics as well as ways to explore engendering trust and fairness in the process of player modeling. 

\section{Acknowledgement}
This work is supported by the National Science Foundation (NSF) under Grant \#1917855. The authors would like to thank all past and current members of the project.

\bibliography{references}

%








\end{document}